\documentclass[twocolumn,showpacs,pra]{revtex4-2}
\usepackage{graphicx,amsmath,mathrsfs, nicefrac}
\usepackage{amssymb}
\usepackage{amsthm,multirow}
\usepackage{bm,bbm}
\usepackage{color}
\usepackage[caption=false]{subfig}

\newcommand{\w}{\omega}

\newcommand{\HMF}{\mathcal{H}_{\rm MF}}
\newcommand{\VBZ}{V_{\rm BZ}}
\newcommand{\JK}{J_{\rm K}}
\newcommand{\Jc}{J_{\rm K,c}}
\newcommand{\TK}{T_{\rm K}}
\newcommand{\nc}{n_c}
\newcommand{\muk}{\mu_{\rm_K}}
\newcommand{\Tcoh}{T_{\rm coh}}

\definecolor{darkgreen}{rgb}{0,0.5,0}
\definecolor{darkblue}{rgb}{0,0,0.5}
\definecolor{purple}{rgb}{0.35,0,0.35}
\definecolor{orange}{rgb}{0.9,0.4,0}

\graphicspath{{./}{./Figures/}}

\DeclareMathAlphabet{\dutchcal}{U}{dutchcal}{m}{n}

\begin{document}
\title{
Kondo screening and coherence in kagome local-moment metals: \\
Energy scales of heavy fermions in the presence of flat bands
}

\author{Christos Kourris}
\author{Matthias Vojta}
\affiliation{Institut f\"ur Theoretische Physik and W\"urzburg-Dresden Cluster of Excellence ct.qmat, Technische Universit\"at Dresden,
01062 Dresden, Germany}

\date{\today}

\begin{abstract}
The formation of a heavy Fermi liquid in metals with local moments is characterized by multiple energy and temperature scales, most prominently the Kondo temperature and the coherence temperature, characterizing the onset of Kondo screening and the emergence of Fermi-liquid coherence, respectively. In the standard setting of a wide conduction band, both scales depend exponentially on the Kondo coupling.
Here we discuss how the presence of flat, i.e., dispersionless, conduction bands modifies this situation. Focussing on the case of the kagome Kondo-lattice model, we utilize a parton mean-field approach to determine the Kondo temperature and the coherence temperature as function of the conduction-band filling $\nc$, both numerically and analytically. For $\nc$ values corresponding to the flat conduction band located at the Fermi level, we show that the exponential is replaced by a linear dependence for the Kondo temperature and a quadratic dependence for the coherence temperature, while a cubic law emerges for the coherence temperature at $\nc$ corresponding to the band edge between the flat and dispersive bands.
We discuss implications of our results for pertinent experimental data.
\end{abstract}

\maketitle


\section{Introduction}

Metals containing local-moment ions herald rich phenomenology thanks to the competition between Kondo screening and multiple types of both symmetry-breaking and topological order \cite{doniach77,stewart01,hvl07,paschen21}. While this phase competition can lead to non-standard forms of quantum criticality, already the physics of the nominally Kondo-screened regime is non-trivial. In particular, the finite-temperature behavior eventually leading to the heavy-fermion metal upon cooling is characterized by multiple crossover temperature scales. In addition to the Kondo temperature $\TK$, the scale below which Kondo screening sets in, the coherence temperature $\Tcoh$ plays an important role, usually defined as the scale below which the system behaves as a coherent Fermi liquid \cite{hewson}. For weak Kondo coupling and a featureless low-energy conduction-electron density of states (DOS), $\TK$ depends exponentially on both the Kondo coupling $\JK$ and the DOS. For $\Tcoh$, early arguments in favor of an even stronger suppression -- dubbed exhaustion -- in particular at small band filling were brought forth \cite{nozieres,nozieres2}, but subsequent work showed that $\Tcoh$ has the same leading exponential behavior as $\TK$ \cite{burdin00,costi02}.

While a large body of heavy-fermion theory is focussed on simple lattices without frustration, various recent developments have brought the interplay of frustration and Kondo physics into focus: a number of heavy-fermion metals with frustrated lattice geometries have been found to show unconventional metallic behavior, such as kagome CeRhSn \cite{tokiwa15,kittaka21}, CePdAl \cite{fritsch14,lucas17,zhao19}, YbAgGe \cite{niklowitz06,tokiwa13}, YbPdAs \cite{xie22}, and pyrochlore Pr$_2$Ir$_2$O$_7$ \cite{nakatsuji06,machida10,pr2ir2o7_grueneisen}. Very recently, Ni$_3$In has been shown to be a correlated kagome metal \cite{ye21}, and Ce$_3$Al has been suggested to follow a similar motif \cite{huang20}.
Finally, moire systems of twisted layers of graphene and transition-metal dichalcogenides have been argued to admit a Kondo-lattice description \cite{song22,kumar22,shi22,sarma23}.
Many of these cases combine Kondo physics with the occurrence of flat (or almost flat) conduction bands. While a few theory studies have appeared addressing flat-band Kondo physics, a systematic study of the relevant crossover scales is missing. It is obvious that the assumption of a wide featureless conduction band is violated here, such that the standard heavy-fermion phenomenology must be modified.

In this paper, we take a step towards closing this gap. For concreteness, we consider a Kondo-lattice model on the kagome lattice, the latter featuring both flat and dispersive tight-binding bands. We utilize a large-$N$ parton mean-field theory to extract both the Kondo and the coherence scale as a function of the Kondo coupling and the conduction-electron filling $\nc$, thus generalizing the work of Ref.~\onlinecite{burdin00} beyond the featureless-DOS case. We indeed find that, due to the flat conduction band, the exponential behavior in both $\TK(\JK)$ and $\Tcoh(\JK)$ is replaced by power laws, being linear for $\TK$ and quadratic for $\Tcoh$. Most strikingly, a filling at the edge between flat and dispersive bands leads to $\Tcoh\propto \JK^3$. While inter-site correlations are not included in our calculation, it is clear that the Fermi-liquid scales determined here are the ones relevant for the competition between local-moment screening and local-moment ordering \cite{doniach77}.
We discuss the relevance to experimental data as well as possible generalizations.

The remainder of the paper is organized as follows:
In Sec.~\ref{sec:model} we introduce the Kondo-lattice model on the kagome lattice and discuss general aspects of its phenomenology.
In Sec.~\ref{sec:parton} we summarize the parton theory, together with the results of Ref.~\onlinecite{burdin00} concerning the Kondo and coherence temperatures.
Our main results are in Sec.~\ref{sec:res} where we show numerical results as well as their analytically extracted weak-coupling behavior for the characteristic energy scales of the kagome Kondo lattice.
An outlook closes the paper. Technical details are relegated to appendices.

We note that the theme of flat-band Kondo physics has appeared in a few earlier papers \cite{tran18,tran19,leelee21,lado21}. The Lieb lattice was studied with a single Kondo impurity (Ref.~\onlinecite{tran18}) and in an Anderson lattice model (Ref.~\onlinecite{tran19}). The Lieb lattice also occurs as effective lattice in the model of Ref.~\onlinecite{lado21}. Ref.~\onlinecite{leelee21} considered a single impurity in a particular flat-band semimetal. None of these papers has studied the Kondo-lattice coherence temperature.


\section{Model and general considerations}
\label{sec:model}

\subsection{Kagome Kondo-lattice model}

\begin{figure}[!tb]
\centering
\includegraphics[width=0.99\columnwidth]{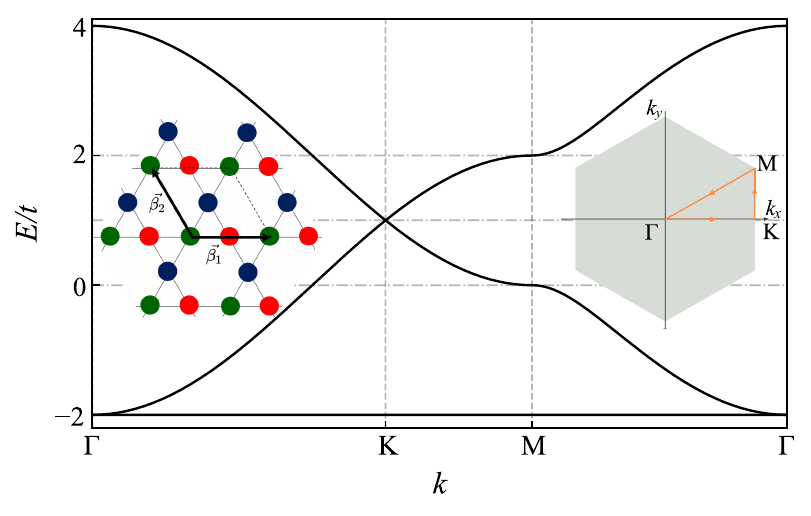}
\caption{
Tight-binding band structure on the kagome lattice, Eq.~\eqref{eq:non_int_bands}, with the flat band located at the bottom \cite{tsignnote}.
The insets show the lattice and the corresponding Brillouin zone.
}
\label{fig:latt}
\end{figure}

As noted above, we are interested in the modifications to standard Kondo phenomenology brought about by flat conduction bands. Many strongly frustrated lattices display flat tight-binding bands as a result of destructive interference. Motivated by its experimental abundance, we choose the kagome lattice, Fig.~\ref{fig:latt}, where we consider a Kondo-lattice model of SU(2) spins $1/2$ (also dubbed $f$ moments) with local Kondo coupling $\JK$. The Hamiltonian reads
\begin{equation}
\label{eq:KLM_hamiltonian}
\mathcal{H} =
t \sum_{\langle ij \rangle}\left ( c^\dagger_{i\sigma}c_{j\sigma} + \rm h.c.\right )
+ \frac{\JK}{2}\sum_i \vec{S}_i \cdot c^\dagger_{i\sigma}\vec{\tau}_{\sigma\sigma'}c_{i\sigma'}
\end{equation}
in standard notation \cite{tsignnote}, with $\langle ij \rangle$ referring to pairs of nearest-neighbor sites, and $\vec\tau$ being the vector of Pauli matrices.

The kagome lattice is a non-Bravais lattice with a unit cell of three sites. The conduction-electron part of Eq.~\eqref{eq:KLM_hamiltonian} can be re-written in Fourier space, yielding the Bloch Hamiltonian matrix
\begin{align}
    \dutchcal{h}(\vec k) &= t\begin{pmatrix}
    0 & 1 + e^{-i\vec k \cdot \vec\beta_1} & 1 + e^{i\vec k\cdot \vec{\beta}_3} \\
1 + e^{i\vec k\cdot \vec\beta_1} & 0 & 1 + e^{-i\vec k\cdot \vec{\beta}_2}\\
1 + e^{-i\vec k\cdot \vec\beta_3} & 1 + e^{i\vec k\cdot \vec{\beta}_2} & 0
\end{pmatrix}
\end{align}
where $\vec k$ is a momentum from the Brillouin zone of the triangular Bravais lattice, the matrix structure refers to sublattice space, and
$\vec{\beta}_1 = (2,0)$, $\vec{\beta}_2 = (-1,\sqrt{3})$, $\vec{\beta}_3 = (-1,-\sqrt{3})$ are the primitive lattice vectors in units of the nearest-neighbor distance $a=1$. Diagonalizing $\dutchcal{h}(\vec{k})$ yields three bands as illustrated in Fig.~\ref{fig:latt}:
\begin{align} \label{eq:non_int_bands}
    \epsilon_{fb} = -2t,~ \epsilon_{\pm} = t \pm t\sqrt{3 + 2g_{\vec k}}
\end{align}
where $g_{\vec k}= \sum_{\ell} \cos( \vec k\cdot \vec{\beta}_{\ell})$. The flat band arises from Wannier states with support on single hexagonal plaquettes which cannot delocalize due to destructive interference, and the flat band touches the dispersing bands \cite{bergman08}. The latter resemble those of the honeycomb lattice, with a Dirac point and two van-Hove singularities (vHs).
The physics of the model will crucially depend on the conduction-band filling, which we define as $\nc = (1/\mathcal{N}) \sum_{i\sigma} c^\dagger_{i\sigma}c_{i\sigma}$, with $\mathcal{N}$ being the number of lattice sites, such that $0\leq\nc\leq 2$.

\begin{figure*}[tb]
\centering
\includegraphics[width=\linewidth]{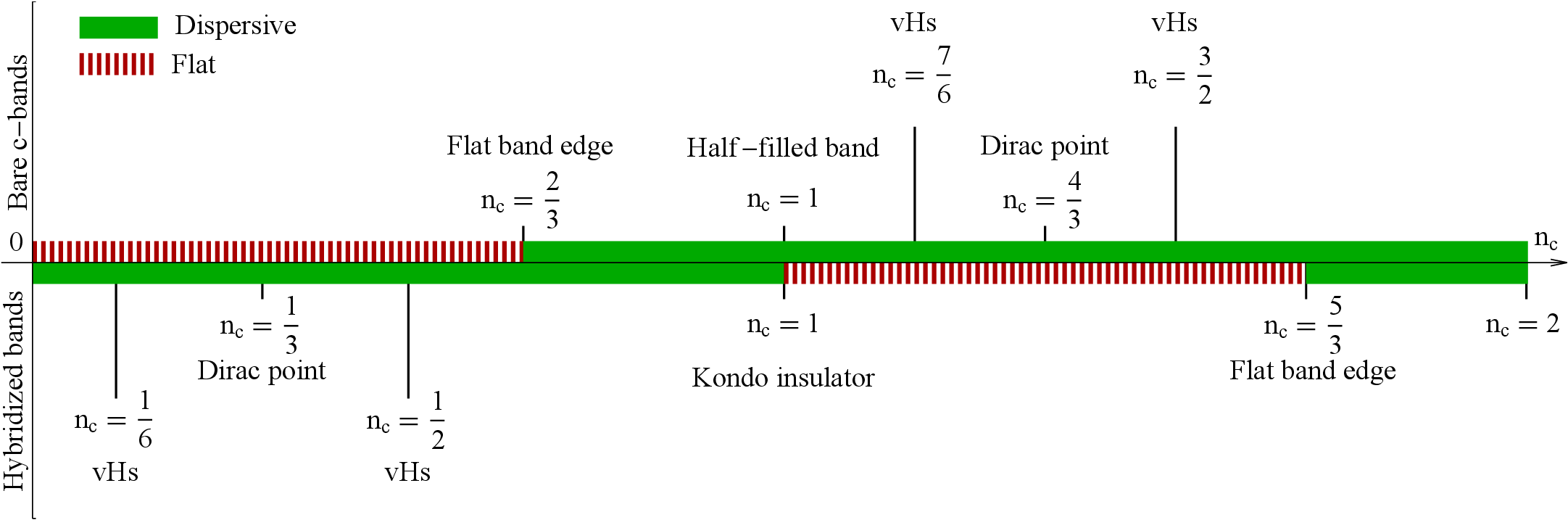}
\caption{
Overview of $T=0$ Fermi-level band-structure features of the kagome Kondo lattice as function of conduction-band filling $\nc$ for the bare conduction band (top) and the hybridized heavy-fermion bands (bottom). Hatched regions correspond to a flat band located at the Fermi level \cite{tsignnote}.
}
\label{fig:fillings}
\end{figure*}

\subsection{Competing phases}

As with any Kondo-lattice model, the physics of Eq.~\eqref{eq:KLM_hamiltonian} is governed by the competition between Kondo screening and inter-moment exchange, mediated by RKKY interactions \cite{doniach77}. In case the latter dominate, magnetically ordered states can arise: a thorough discussion of those in the limit of classical spins (where Kondo screening is suppressed) has been given in Refs.~\onlinecite{batista14,ghosh16}; they include coplanar $\sqrt{3}\times\sqrt{3}$ as well as various noncoplanar orders, including multi-$Q$ states. For quantum spins $1/2$, it is conceivable that fractionalized Fermi-liquid states (FL$^\ast$) occur, where the $f$ moments do not order, but realize a fractionalized spin liquid \cite{senthil03,senthil04}.

In this work, we will focus on the opposite case when Kondo screening dominates. For this, we ignore the effect of inter-site interactions and treat the local moments as quantum spins. With the local moments being screened, a paramagnetic Fermi-liquid phase emerges whose characteristic energy scales -- $\TK$ and $\Tcoh$ \cite{hewson} -- will, however, be strongly influenced by the presence of flat bands. We note that the heavy Fermi liquid is potentially unstable to superconductivity (with the tendency being enhanced for a flat heavy band \cite{kopnin11,huber18}); we will ignore such an instability.

We note that Kondo screening is inevitably realized in the strong-coupling limit $\JK\gg t$. Here, the ground state is a product of local $c$-$f$ singlets, with small perturbative corrections, at $\nc=1$. Departing from $\nc=1$, unpaired carriers will move in this singlet background forming a standard (heavy) Fermi liquid.

\subsection{Renormalized band structure}

It is instructive to discuss the heavy Fermi liquid in a picture of renormalized bands: we recall that the relevant low-energy and low-temperature physics is captured by hybridized $c$ and $f$ bands \cite{hewson}. Assuming non-dispersive bare $f$ electrons as in Eq.~\eqref{eq:KLM_hamiltonian}, the hybridized band structure will consist of two triplets of bands, separated by a hybridization gap, with both triplets featuring one flat and two dispersive bands. Moreover, the heavy Fermi liquid will obey Luttinger's theorem, i.e., its Fermi volume equals $(\VBZ/2)[3(\nc+1)\,{\rm mod}\,2]$ where $\VBZ$ is the momentum-space volume of the first Brillouin zone, and the factors of two and three account for spin and the number of sites per cell, respectively. In contrast, the behavior at elevated temperatures above $\TK$ is governed by the unhybridized conduction electrons which are weakly scattered off the local moments.

These considerations enable us to map the special band fillings of the problem at hand, as shown in Fig.~\ref{fig:fillings}. The primary focus of this paper will be on fillings $0<\nc\lesssim2/3$ where the bare conduction band is flat, but the hybridized band structure features dispersive bands at the Fermi level, such that coherent transport can be expected. A detailed study of the regime $1<\nc\leq5/3$, with a flat heavy-fermion band at the Fermi level (i.e. with ultra-heavy fermions), is left for future work. This latter regime is apparently prone to further flat-band-driven ordering tendencies, driven by a combination of RKKY interactions and effects of quantum geometry \cite{mera22,bernevig22}.


\section{Parton theory}
\label{sec:parton}

To describe the heavy Fermi liquid, we utilize the standard Abrikosov fermion representation for the local moments, $\vec{S}_i = f^\dagger_{i\sigma}\vec{\tau}_{\sigma\sigma'} f_{i\sigma'} / 2$, together with a slave-boson approximation of the Kondo interaction. This approach has been used in numerous earlier works \cite{coleman83,coleman84,coleman85,burdin00,hewson,senthil03,senthil04}, and we quickly summarize its main ingredients.

\subsection{Mean-field approximation}

Inserting the above spin representation into $\mathcal{H}$ yields a quartic Kondo term which is decoupled in the singlet particle--hole channel using a bosonic field $r_i$ at lattice site $i$. The key approximation is to then assume that $r_i$ is static, corresponding to a saddle-point approximation which becomes exact in the large-$N$ limit where the spin symmetry is generalized to SU($N$) for both local moments and bath electrons. Consistent with this, the Hilbert-space constraint for the $f$ fermions, $\sum_\sigma f_{i\sigma}^\dagger f_{i\sigma}=1$, is enforced only on average via Lagrange multipliers $\lambda_i$. Assuming furthermore a homogeneous saddle-point solution, $r_i\to r$ and $\lambda_i\to \lambda$, and introducing a chemical potential $\mu$ for the conduction electrons, the Hamiltonian takes the form (up to constants)
\begin{align}
\label{eq:hamiltonian_interacting}
    \HMF = &- t \sum_{\langle ij \rangle}\left ( c^\dagger_{i\sigma}c_{j\sigma} + \rm h.c.\right ) - r\sum_{i} \left ( f^\dagger_{i\sigma} c_{i\sigma}+ \rm h.c. \right ) \notag\\
    &-\lambda \sum_{i} f_{i\sigma}^{\dagger}f_{i\sigma}-\mu \sum_{i} c_{i\sigma}^{\dagger}c_{i\sigma},
\end{align}
together with the self-consistency condition
\begin{align} \label{eq:mean_field_parameter}
r = \frac{\JK}{2\mathcal{N}}\sum_{i\sigma} \langle f^{\dagger}_{i\sigma}c_{i\sigma}\rangle.
\end{align}
The mean-field parameter $r$ encodes the emergent hybridization between $c$ and $f$ bands and is thus a measure of Kondo screening. At high temperatures, the mean-field equation is solved by $r=0$ only, whereas a finite-$r$ solution emerges at low $T$. All solutions preserve SU(2) spin symmetry such that all bands and expectation values are spin-degenerate; therefore we will drop the spin index from now on.
The mean-field problem can be solved in Fourier space. Using the spinor
\begin{equation}
    \psi_{\vec k} = (c^A_{\vec k}, c^B_{\vec k}, c^C_{\vec k}, f^A_{\vec k}, f^B_{\vec k}, f^C_{\vec k})
\end{equation}
we re-write the Hamiltonian per spin flavor as
\begin{align}
    \mathcal{H} = \sum_{\vec{k}} \psi^\dagger_{\vec{k}} \dutchcal{H}(\vec{k}) \psi_{\vec{k}} + \mathcal{N} \left( \frac{r^2}{\JK} + \frac{\lambda}{2} + \frac{\mu\nc}{2}\right)
\end{align}
where the $6\times 6$ Bloch matrix $\dutchcal{H}(\vec{k})$ is given by
\begin{align}
       \dutchcal{H}(\vec{k}) = \left(\begin{array}{c|c}
    	\dutchcal{h}(\vec{k}) - \mu \mathbb{I} & -r\mathbb{I}\\
    	\hline
    	-r\mathbb{I} & - \lambda \mathbb{I}
    \end{array}\right).
\end{align}
The resulting band structure consists of six hybridized bands, obtained by diagonalizing $\dutchcal{H}$,
\begin{align}
    \bar E_\text{p}^{q} &=  - \mu + \frac{1}{2} \left ( \epsilon_{p} - \lambda' + q \sqrt{(\epsilon_{p} + \lambda')^2 + 4r^2} \right )\label{eq:six_bands}
\end{align}
where $q = \pm$ denotes the upper and lower triplet of bands, $p\in \{fb,+,-\}$, and $\lambda'=\lambda-\mu$. As noted above, the band structure is split into two triplet of bands, separated by a gap. Both triplets support a flat band, a Dirac point, located at $\bar E^{+}_{-} = \bar E^{+}_{+}$ and $\bar E^{-}_{-} = \bar E^{-}_{+}$ respectively, and two vHs each.

The mean-field free-energy density per spin flavor can be written as
\begin{align}\label{eq:free_energy}
\begin{split}
    \mathcal{F}(T) = &- T\int_{-\infty}^{\infty}d\epsilon\ \rho_0(\epsilon) \sum_{p=\pm} \ln \left (1 + e^{-\beta \bar E^{p}(\epsilon)} \right )\\
    &+ \left( \frac{r^2}{\JK} + \frac{\lambda}{2} + \frac{\mu\nc}{2}\right)
    \end{split}
\end{align}
where $\rho_0(\epsilon)$ is the DOS of the (free) conduction electrons, and $\bar E^{\pm} = -\mu + \frac{1}{2}\left (\epsilon - \lambda' \pm \sqrt{(\epsilon + \lambda')^2 + 4r^2} \right )$. Minimizing $\mathcal{F}(T)$ with respect to the three parameters $r, \mu, \lambda$ leads to the mean-field equations \cite{burdin00}:
\begin{align}
 -\frac{1}{\JK} &= \int_{-\infty}^{\infty}\frac{d\epsilon}{\epsilon + \lambda}\ \rho_0\left ( \epsilon + \mu - \frac{r^2}{\epsilon +   \lambda}\right )n_F(\epsilon), \label{eq:MF_1}\\
    \frac{1}{2} &= \int_{-\infty}^{\infty} \frac{d\epsilon \, r^2}{(\epsilon + \lambda)^2}\ \rho_0\left ( \epsilon + \mu - \frac{r^2}{\epsilon +  \lambda}\right )n_F(\epsilon), \label{eq:MF_2}\\
    \frac{n_c}{2} &= \int_{-\infty}^{\infty}\ d\epsilon\ \rho_0\left ( \epsilon + \mu - \frac{r^2}{\epsilon +  \lambda}\right )n_F(\epsilon) \label{eq:MF_3}
\end{align}
with $n_F(\epsilon)$ the Fermi function. In the remainder of the paper, we will solve these equations numerically for arbitrary $\JK$ as well as analytically in the small-$\JK$ limit.


\subsection{Kondo temperature}

The Kondo temperature $\TK$ is a crossover temperature which is usually associated with Kondo screening becoming substantial upon cooling. Different quantitative definitions are routinely employed, ranging from elevated-temperature fits of the spin susceptibility to $\chi(T)=C/(T+\sqrt{2}\TK)$ \cite{wilson75} to the renormalized dimensionless Kondo coupling becoming unity within a renormalization-group framework. The present parton mean-field theory displays a continuous phase transition as function of $T$ where $r$ becomes non-zero indicating the onset of screening; the temperature of this transition is taken as the Kondo temperature $\TK$. It is known that this transition is an artifact of the mean-field approximation and turns into a crossover once fluctuations of the compact U(1) gauge field are taken into account \cite{compact}.

Solving the MF equation~\eqref{eq:MF_1} for the onset of a finite $r$ yields the equation for $\TK$:
\begin{align}\label{eq:kondo_final_MF}
    \frac{2}{\JK} = \int_{-\infty}^{\infty}d\epsilon\ \frac{\rho_0(\epsilon + \mu_0)}{\epsilon}\tanh\left ( \frac{\epsilon}{2\TK}\right ) .
\end{align}
with $\mu_0$ the bare conduction-electron chemical potential corresponding to $n_c$. We note that, within the mean-field approximation, $\TK$ for the Kondo lattice is identical to that of the single-impurity problem on the same lattice, as only local properties of the conduction band enter. For a flat conduction band with DOS $\rho$, the mean-field theory delivers $\TK \propto e^{-1/(\rho\JK)}$; this exponential agrees with the exact solution of the single-impurity problem \cite{hewson,tknote}.


\subsection{Coherence temperature}

The coherence temperature $\Tcoh$ marks the crossover temperature below which the heavy-electron state behaves as a coherent Fermi liquid. Experimentally, it is often defined as the temperature location of the maximum in the resistivity. An alternative definition, more suitable for the present mean-field theory, is via the effective bandwidth of the heavy quasiparticles which can be efficiently obtained from DOS of the hybridized bands at the Fermi energy, $\Tcoh=1/\rho(\w=0)$, evaluated at $T=0$. For $r\ll D$ this can be approximated by $\Tcoh = r^2/D$, for $D$ the conduction-electron bandwidth and assuming a featureless hybridized DOS. We will employ this expression for $\Tcoh$ throughout the paper, note its limitations when appropriate, and also offer an a-posteriori justification by a direct comparison with the mean-field results for $1/\rho(\w=0)$ in Appendix~\ref{apx:Tcoh_solutions}.

Determining $\Tcoh=r^2/D$ requires solving the MF equations \eqref{eq:MF_1}-\eqref{eq:MF_3} in the $T\to 0$ limit where we can then replace $n_F(\epsilon) \to 1$ when using the integration limits $(\epsilon_L, 0)$, where $\epsilon_L = -\mu - \frac{\lambda'}{2} - \sqrt{\left( \frac{\lambda'}{2}\right)^2 + r^2 }$, and $\mu$ corresponds to the $c$-electron chemical potential which differs from ``bare'' value $\mu_0$.
We recall that $\lambda'=\lambda-\mu$; in the weak-coupling limit, and by making use of Luttinger's theorem, we find (see Appendix~\ref{apx:chem_pot}) that $\lambda \propto r^2/D$.


\begin{figure*}[tb]
  \centering
  \includegraphics[width=0.96\textwidth]{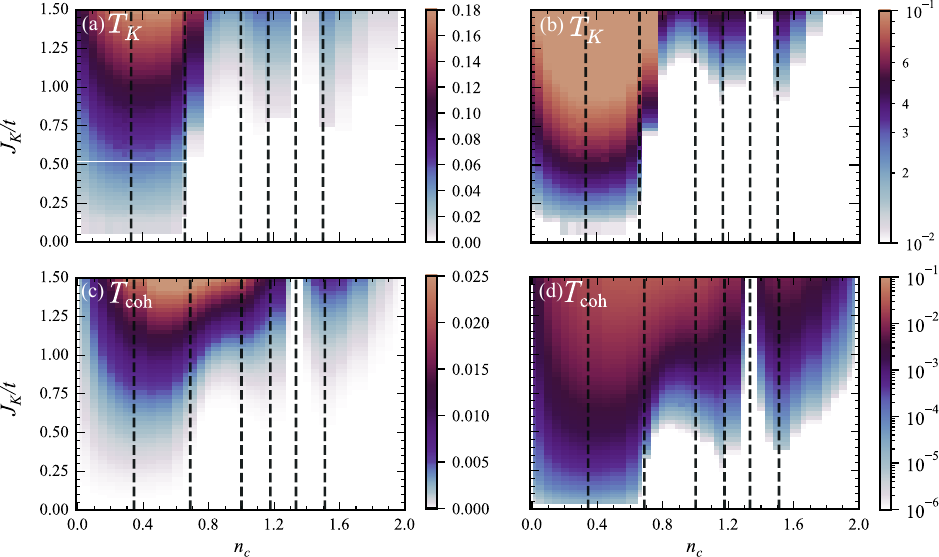}
\caption{(a,b) Kondo temperature and (c,d) coherence temperature as function of conduction-band filling $\nc$ and Kondo coupling $\JK$, shown on (a,c) linear and (b,d) logarithmic color scales. The vertical lines correspond to the special fillings $\nc = \nicefrac{1}{3},\nicefrac{2}{3},1,\nicefrac{7}{6},\nicefrac{4}{3}$ and $\nicefrac{3}{2}$.
Screening is seen to be enhanced in the regime of the flat conduction band, $0<\nc<2/3$.
}
  \label{fig:heatmaps}
\end{figure*}

\section{Results: Numerical and analytical}
\label{sec:res}

In this section we present our numerical and analytical results for Kondo and coherence temperatures for ranges of the conduction-band filling $\nc$, obtained from solving the mean-field equations of Sec.~\ref{sec:parton}. Details of the analysis can be found in the appendices.


\subsection{Overview of numerical results}

We have solved the MF equations \eqref{eq:MF_1}-\eqref{eq:MF_3} at finite $T$ using a momentum grid of $180^2$ points. $\Tcoh$ is obtained as $r^2/D$, with $D=6t$ at $T=0$, while $\TK$ is extracted by determining $r(T)$ for a series of temperatures which is then fit to a form $r = A\sqrt{T-\TK}$ for $T$ near $\TK$, see Appendix~\ref{apx:TKondo_solutions} for a sample fit. Finite-size effects limit the accessible Kondo temperatures to $\TK\gtrsim 10^{-2}t$.

Fig.~\ref{fig:heatmaps} shows Kondo and coherence temperatures as function of both band filling and $\JK$. As expected from the general dependence of these temperatures on the Fermi-level DOS of the conduction electrons, both $\Tcoh$ and $\TK$ are enhanced for fillings $\nc<2/3$, corresponding to a flat conduction band, compared to the ``generic'' situation for $\nc>2/3$. We also see a moderate enhancement for the van-Hove fillings $7/6$ and $3/2$, while both scales are strongly suppressed for the Dirac-point filling $4/3$. In the following, we will study the flat-band situation in more detail.


\subsection{Weak-coupling expansion}

To access the asymptotic low-energy behavior, we analyze the mean-field solutions at weak Kondo coupling, $\JK\ll t$. We focus on the filling regime dominated by the flat conduction band, i.e., $n_c\lesssim 2/3$; for generic fillings above the flat band the results of Ref.~\onlinecite{burdin00} apply.

The behavior of the mean-field solutions is determined by the conduction-electron DOS (per spin). Its exact form can be expressed as
\begin{align}
 \rho_0(\epsilon) = \frac{1}{3}\delta({\epsilon}) + \frac{2}{3}|\epsilon - 3| \rho_\Delta[ (\epsilon - 3)^2 - 3]
\end{align}
where $\rho_\Delta$ is the DOS of the triangular lattice \cite{kogan21}, and we have shifted the bottom of the band, identical to the flat-band energy, to $\epsilon = 0$. Being interested in $\nc\lesssim 2/3$, we approximate this DOS as
\begin{align}
\label{eq:dos_approx}
\rho_0(\epsilon) \approx \frac{1}{3}\delta(\epsilon) + \tilde \rho \,\Theta(\epsilon) \Theta(\tilde D-\epsilon)
\end{align}
with $\tilde \rho = \frac{1}{2\sqrt{3}\pi}\frac{1}{t}$ representing the dispersive-band contribution and $\tilde D = \frac{4\pi}{\sqrt{3}}t$ the resulting band width. This approximation neglects the Dirac point and the vHs.

In order to determine $\TK$, it is apparent from Eq.~\eqref{eq:kondo_final_MF} that the functional form of the conduction-electron chemical potential at $\TK$ is necessary. In the weak-coupling limit, only the asymptotic form of $\mu_0(T\to 0)$ is required, which is determined by
\begin{align} \label{eq:filling_kondo_mu}
\frac{n_c}{2} = \int_{-\infty}^{\infty} d\epsilon\ \rho_0(\epsilon) n_F(\epsilon - \mu_0).
\end{align}
Similarly, and following from the approximations already made, $\Tcoh$ is determined by solving Eq.~\eqref{eq:MF_1}-\eqref{eq:MF_3} self-consistently, and thus by first determining the dependence of the chemical potentials $\mu$ and $\lambda'$ on $r$. The details of the calculation can be found in Appendix \ref{apx:chem_pot}.

In the remainder of this section, we describe the analytical results for different filling ranges vis-a-vis the numerical data.


\subsection{$0<n_c<2/3$: Partially filled flat conduction band}

\subsubsection{Kondo temperature}

The flat band implies $\mu_0(T\!=\!0) = 0$ for all $0<\nc\leq 2/3$. To leading order in $T$, we find $\mu_0(T) = \alpha(n_c) T$ with the coefficient
\begin{align}
\label{muprefac}
 \alpha(n_c) = \ln \left ( \frac{3n_c}{2-3n_c}\right )\,.
\end{align}
The divergence of $\alpha(n_c)$ as $n_c\to 2/3$ implies that the case of a filled flat band needs to be treated separately. Upon substitution in Eq.~\eqref{eq:kondo_final_MF}, we can extract the asymptotic behavior at low $\TK$ as
\begin{align}
\label{lintk}
    \TK = \frac{1}{2}m(n_c) \JK + \mathcal{O} (\JK^2\ln \JK)
\end{align}
where $m(n_c) = \frac{1}{3} \frac{\tanh(\alpha(n_c)/2)}{\alpha(n_c)}$. The prefactor is symmetric about $n_c = \frac{1}{3}$, signalling an emergent particle-hole symmetry within the flat band \cite{prefactor_comment}.

\begin{figure}[tb]
 \centering
   \includegraphics[width=\columnwidth]{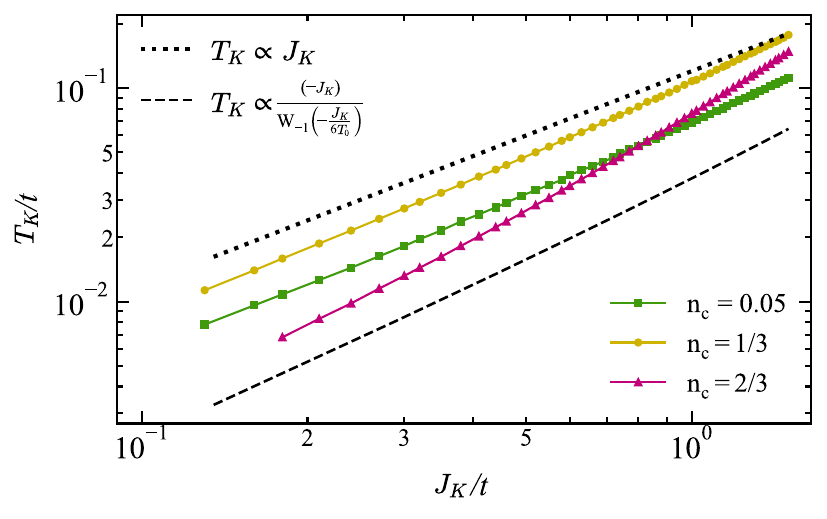}
 \caption{
 Kondo temperature $\TK$ vs. Kondo coupling $\JK$ for different fillings $n_c$.
 The dotted (dashed) line indicates a linear (log-linear, Eq.~\eqref{eqloglinear}) dependence for reference.
 }
\label{fig:kondo_numerical}
\end{figure}

Fig.~\ref{fig:kondo_numerical} illustrates that the numerical data indeed follow the linear $\JK$ dependence of Eq.~\eqref{lintk} over a wide range of $\JK$ except very close to $\nc=2/3$ where the proximate dispersive band influences $\TK$ already at small $\JK$. A detailed comparison, including the subleading corrections obtained analytically, can be found in Appendix~\ref{apx:TKondo_solutions}.

\subsubsection{Coherence temperature}

\begin{figure}[tb]
 \centering
   \includegraphics[width=\columnwidth]{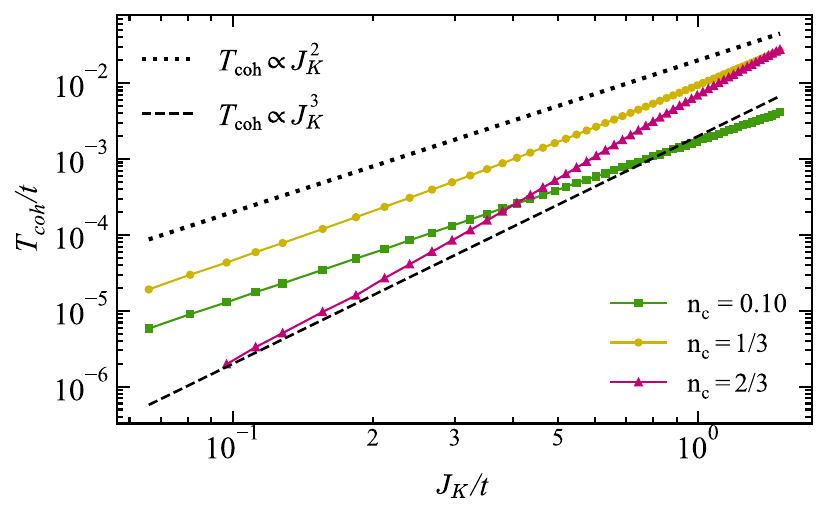}
 \caption{
 Coherence temperature $\Tcoh$ vs. Kondo coupling $\JK$ for different fillings $n_c$.
 The dotted (dashed) lines indicate a quadratic (cubic) dependence, $\Tcoh\propto\JK^2$ ($\Tcoh\propto\JK^3$), respectively.
 }
 \label{fig:logTcoh_logJk}
\end{figure}

At $T=0$, we find that the mean-field parameter $r\propto\JK$. Therefore, the coherence temperature is quadratic in $\JK$,
\begin{align}\label{eq:anal_coh}
    \Tcoh = \frac{1}{2}g(n_c) \frac{\JK^2}{D} + \mathcal{O}(\JK^3\ln \JK)
\end{align}
where $g(n_c) = \frac{1}{2} n_c ( \frac{2}{3} - n_c )$, with the prefactor again symmetric around $n_c = \frac{1}{3}$. The numerical data shown in Fig.~\ref{fig:logTcoh_logJk} obey this quadratic behavior over a wide range of $\JK$ except near $\nc=2/3$. We refer the reader to Appendix~\ref{apx:Tcoh_solutions} for details on the calculation and results for the subleading corrections.

\subsection{$n_c=2/3$: Filled flat conduction band}

\subsubsection{Kondo temperature}

The divergence of the linear prefactor of the chemical potential, Eq.~\eqref{muprefac}, signifies a qualitative change once the flat band is filled. As is sketched in Appendix~\ref{apx:chem_pot}, the chemical potential is now $\mu(T) = T\ln\left (\frac{T_0}{T} \right )$ with $T_0 \simeq 3.126t$. The Kondo MF equation reduces to
\begin{align}\label{eq:kondo_2over3}
    \JK = 6\TK \ln\left ( \frac{T_0}{\TK} \right ) \cdot \left (1 - 3a\tilde \rho \TK  \ln \frac{T_0}{\TK} \right )\,.
\end{align}
The leading part of this equation can be inverted using the Lambert $W$ function:
\begin{equation}
\TK = - \frac{\JK}{6 W_{-1}(-\JK/(6T_0))}\,.
\end{equation}
Using the asymptotic expansion of $W_{-1}$ yields
\begin{equation}
\TK \approx \frac{\JK}{6 \ln(6T_0/\JK)}\,.
\label{eqloglinear}
\end{equation}

\subsubsection{Coherence temperature}

At $\nc=2/3$ the coherence behavior is also qualitatively different, as indicated by the singularity in Eq.~\eqref{eq:anal_coh}. The chemical potential $\lambda'$, which for fillings $1/3 < n_c < 2/3$ is such that $\lambda' < 0$, is now related to the mean-field parameter $r$ through $r\approx \frac{3^{1/4}}{\sqrt{2\pi}}|\lambda'|^{3/2}$. Correspondingly, the mean-field parameter $r$ attains a power-law dependence, $r\propto\JK^{3/2}$, leading to
\begin{align}\label{eq:coh_temp_2over3}
 \Tcoh = \frac{1}{3^{3/2}\pi} \frac{\JK^{3}}{D^2}
\end{align}
using $D=6t$, in agreement with the numerical data in Fig.~\ref{fig:logTcoh_logJk}.

The analysis makes clear that $\nc = 2/3$ is special because of the touching of a flat and a quadratically dispersing conduction band. This band touching is symmetry-protected so that, in the absence of an additional spin-orbit coupling, the two bands will always touch \cite{bergman08,kang20}. In the opposite case of a gap between a flat and a dispersing band, the asymptotic behavior of both $\TK$ and $\Tcoh$ would be perfectly symmetric with respect to the half-filled flat band, without a special power law at the edge.


\subsection{$2/3<n_c<2$: Dispersive conduction band}

\subsubsection{Kondo temperature}

For fillings above the flat band, and away from the fillings which correspond to the Dirac point and the two vHs, the system is a normal metal with a finite DOS. This case has been considered fully in Ref.~\onlinecite{burdin00} where full expressions with exact prefactors for both the Kondo and coherence temperature are provided. For the fillings considered here, $\mu(T\to 0)$ is a filling-dependent constant, for which the Kondo MF equation reproduces the conventional behavior,
\begin{align}
 \TK = F_K(\nc) \exp\left (-\frac{1}{\tilde \rho \JK} \right );
\end{align}
the prefactor $F_K$ is a smooth function of $\nc$ and has been calculated for a flat conduction band in Ref.~\onlinecite{burdin00}. Such exponential behavior is consistent with our numerical data, where $\tilde\rho$ needs to be replaced with the actual value of the conduction-electron DOS at $\nc$.

\begin{figure}[tb]
 \centering
   \includegraphics[width=\columnwidth]{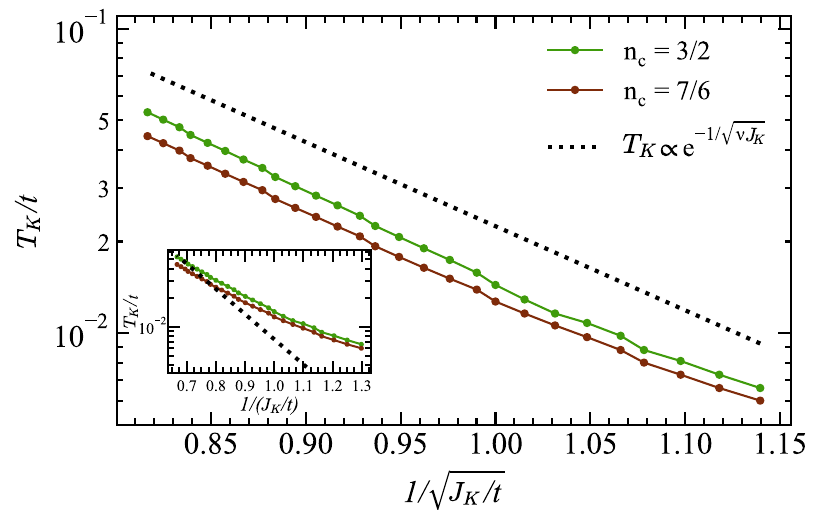}
 \caption{
 Kondo temperature $\TK$ vs. Kondo coupling $\JK$ for the fillings corresponding to the vHs. The dotted line indicates the exponential dependence, $\TK \propto \exp(-1/\sqrt{\nu \JK})$, with $\nu = 0.024$. The inset illustrates the failure of the standard exponential fitting.
 }
 \label{fig:Tk_vhs}
\end{figure}

For the special fillings corresponding to the vHs and the Dirac point, the featureless-DOS approximation is invalid. Previous work on the Kondo effect on the square lattice found that the logarithmic vHs yields a Kondo temperature scaling as  $\TK^{\rm VH} \sim \exp(-1/\sqrt{\nu \JK})$, for $\nu$ proportional to the inverse bandwidth \cite{gogolin93,irkhin11,zhuravlev2018}. The data for the Kondo temperature at the vHs, depicted in Fig.~\ref{fig:Tk_vhs}, is in agreement with this scaling.

For the Dirac point, the absence of states at the Fermi level causes the Kondo temperature to vanish exactly at small $\JK$, in agreement with the results of the pseudogap Kondo problem \cite{withoff90,GBI98,VF04,FV04,principi15}. Screening sets in for $\JK>\Jc$, with a quantum phase transition at a critical coupling $\Jc \approx 1.7t$.

\subsubsection{Coherence temperature}

For the flat DOS that corresponds to the fillings above the flat band, our calculation reduces to a simplified version of the calculation in \cite{burdin00}, with a flat DOS. We find that the exponential dependence of the coherence temperature is recovered,
\begin{equation}
 \Tcoh = F_{\rm coh}(\nc) \exp \left ( -\frac{1}{\tilde \rho \JK} \right ) .
\end{equation}
The prefactor is again a smooth function of $\nc$ \cite{burdin00}. To match the numerical data, $\tilde\rho$ again needs to be replaced by $\rho_0(\mu_0)$ corresponding to the actual $\nc$.

\begin{figure}[tb]
 \centering
   \includegraphics[width=\columnwidth]{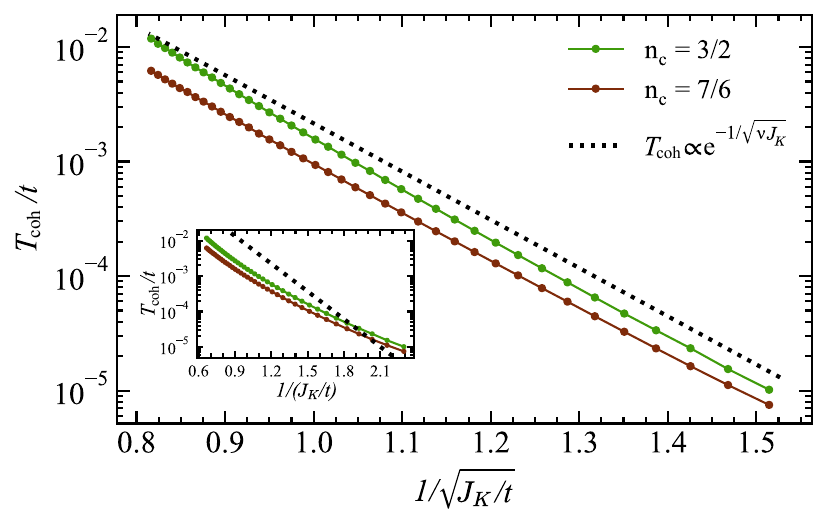}
 \caption{
 Same as Fig.~\ref{fig:Tk_vhs}, but now for the coherence temperature $\Tcoh$, with the fit parameter $\nu = 0.011$.
 }
 \label{fig:Tcoh_vhs}
\end{figure}

As above, the vHs and the Dirac point are beyond this approximation. The data corresponding to the coherence temperature at the vHs are depicted in Fig.~\ref{fig:Tcoh_vhs}. To our knowledge, the coherence temperature has not been studied analytically for the case of a logarithmic vHs. The same applies to the Dirac point; we note that the present kagome-lattice case yields a metallic state at this filling, $\nc=4/3$, in contrast to the honeycomb-lattice (i.e. graphene) case where $\nc=1$ at the Dirac point, thus resulting in a Kondo insulator. We leave a detailed studied of these special fillings for future work.

\subsection{Thermodynamics}

An essential outcome of the parton theory is that all the physical properties are determined by the structure of the hybridized bands. In the following, we briefly discuss the thermodynamics for band fillings corresponding to the flat conduction band.

As illustrated in Fig.~\ref{fig:fillings}, fillings $\nc \lesssim 2/3$ yield dispersive (instead of flat) heavy fermion bands. Hence, the asymptotic low-$T$ properties are non-singular, i.e., that of a standard heavy Fermi liquid, with heat capacity $C\sim\gamma T$ etc. This behavior can be expected for $T\ll\Tcoh$ whereas the specific features of the kagome-lattice band structure will only enter at elevated temperatures.

\begin{figure}[!bt]
\centering
\includegraphics[width=\columnwidth]{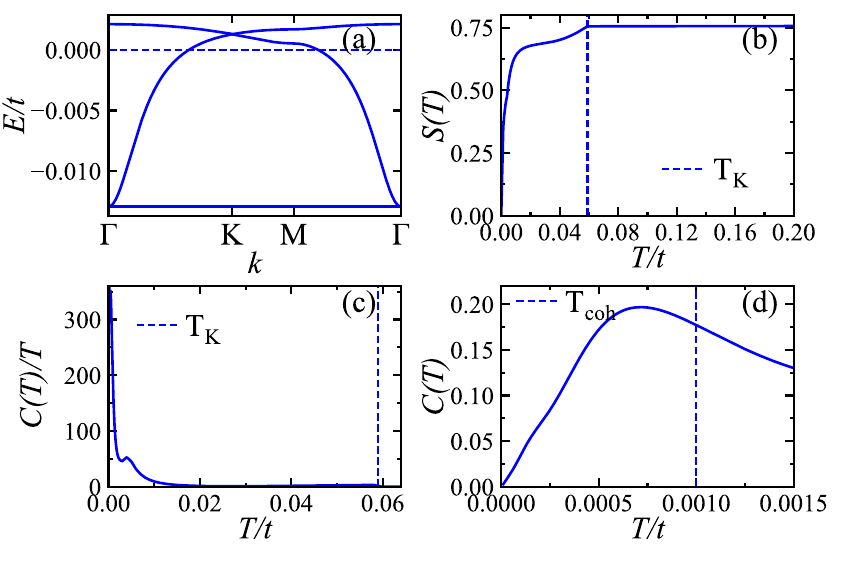}
\caption{
Mean-field results for $\JK = 0.97t$ and $\nc = 0.03$ where $\TK\approx0.058t$ and $\Tcoh\approx0.001t$.
(a) Parton band structure at $T=0$.
(b) Entropy per spin, $S(T)$.
(c,d) Heat capacity, plotted as $C(T)/T$ up to $\TK$ and $C(T)$ at very low $T$, respectively.
}
\label{fig:observables}
\end{figure}

We have used the parton theory to compute the entropy density per spin, $S(T)$, and the resulting heat capacity, $C = TdS/dT$. A sample result for $\JK = 0.97t$ and $\nc = 0.03$ is shown in Fig.~\ref{fig:observables}. At this filling, the $T=0$ Fermi level is below the Dirac point of the heavy bands. At $T\ll\Tcoh$ both $S$ and $C$ depend linearly on $T$, with a strongly enhanced specific-heat coefficient $C/T$. Upon heating up to $\Tcoh$, $C/T$ significantly decreases due to the proximity to the heavy-electron Dirac point. For $T>\TK$ the mean-field theory delivers $r=0$, such that $S$ consists of a small and weakly $T$-dependent $c$-electron contribution and that from a localized $f$ level at the Fermi energy, resulting in a value slightly larger than $\ln 2$. The dominant release of entropy upon cooling happens immediately below $\TK$ where the originally flat $c$ band is removed from the Fermi level.

\section{Summary and outlook}
\label{sec:concl}

Using a parton mean-field theory, we have studied the energy/temperature scales characterizing a heavy Fermi liquid in the kagome Kondo lattice model, with focus on electronic fillings corresponding to the flat conduction band and its vicinity. In accordance with the enhanced density of states, we have found a strong enhancement of both the Kondo and coherence temperatures which display power-law instead of exponential dependencies on the Kondo coupling. Such behavior is expected to be generic for flat-band Kondo-lattice systems and to also apply in three space dimensions, e.g., for pyrochlore-lattice systems with flat conduction bands.
We note that the Lieb lattice treated in Refs.~\onlinecite{tran18,tran19} features inequivalent sites and hence displays multiple Kondo scales.

Beyond the single-site approximation employed here, various instabilities to symmetry-broken as well as topological phases may occur, driven by RKKY-type interactions \cite{doniach77,pines08}. On the kagome lattice, these will be influenced in a non-trivial fashion by sublattice interference \cite{thomale,batista14,ghosh16} and effects of quantum geometry \cite{mera22,bernevig22}. The present work establishes the screening scales against which other phases need to compete; detailed studies of this competition are left for future work.

Flat electronic bands are a recurring theme in many experimental contexts, one being magic-angle Moire systems \cite{macdonald20,song22,kumar22,shi22,sarma23}. In most cases, the bands are not exactly but only approximately flat. This also applies to many of the heavy-fermion compounds mentioned in the introduction whose kagome structure is distorted. Whether or not Kondo enhancement by flat conduction bands is relevant there is unclear at present; density functional studies of the band structure assuming localized $f$ electrons might be helpful. Adapting the analysis of the present paper to almost flat bands is the subject on ongoing work. Our results also call for numerical studies beyond the static mean-field approximation to understand the dynamic properties in the flat-band-dominated regime.

\acknowledgments
We thank P. Consoli and L. Janssen for discussions and collaborations on related work.
Financial support from the Deutsche Forschungsgemeinschaft through SFB 1143 (Project-id No. 247310070, No. 390858490 ) and the W\"urzburg-Dresden Cluster of Excellence on Complexity and Topology in Quantum Matter -- \textit{ct.qmat} (EXC 2147, project-id 390858490) is gratefully acknowledged.


\appendix

\section{Chemical potential analysis}
\label{apx:chem_pot}


\subsection{Filling dependence of $\mu_0(n_c)$ at low temperatures}

\begin{figure}[tb]
\centering
\includegraphics[width=\columnwidth]{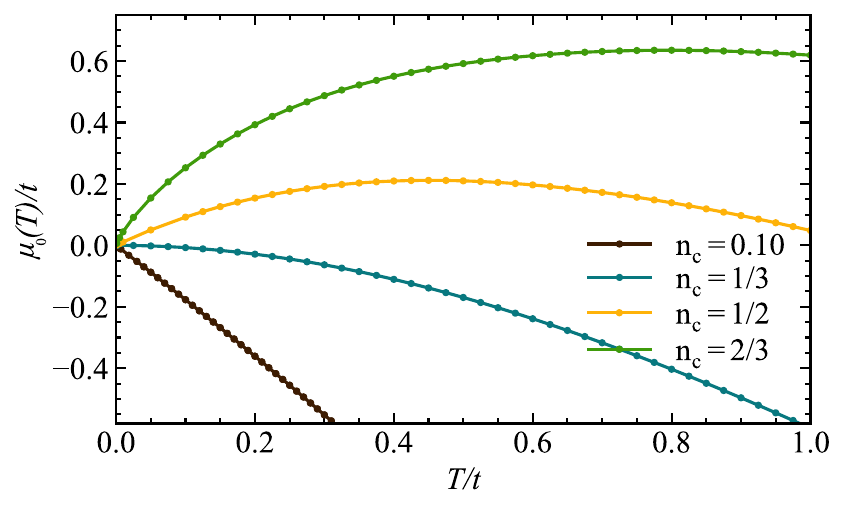}
\caption{
Numerically determined conduction-electron chemical potential $\mu_0(T)$ for different fillings $n_c$ inside the kagome-lattice flat band.
}
\label{fig:non_int_chem_pots}
\end{figure}

The approximate form of the non-interacting DOS, Eq.~\eqref{eq:dos_approx}, may be used to obtain an asymptotic form of the ``bare'' chemical potential as $T\to 0$. The filling equation, $n_c/2 = \int_{-\infty}^{\infty} d\epsilon\ \rho_0(\epsilon) n_F(\epsilon - \mu_0)$, reduces to
\begin{align}
   \nc = \frac{2}{3}\frac{1}{1 + e^{-\mu_0/T}} + \frac{1}{\sqrt{3}\pi} \frac{T}{t}\ln{\left (\frac{e^{\mu_0/T} + 1}{e^{- (\tilde D - \mu_0 )/T} + 1}\right )}\label{eq:filling_mu}
\end{align}
The chemical potential for a series of fillings inside the flat band is presented in Fig.~\ref{fig:non_int_chem_pots} and analytically obtained in the following subsections.

\subsubsection{$0<n_c<2/3$}

In the case of a partially filled flat band, the asymptotic behavior is well captured by the first term of Eq.~\eqref{eq:filling_mu}. The solution is given by,
\begin{align}\label{eq:mu0_partially_filled}
 \mu_0(T) = T \ln \left ( \frac{3n_c}{2-3n_c}\right )
\end{align}
We see that the prefactor changes sign at $n_c = \frac{1}{3}$, where the flat band is half filled and diverges both for $\nc\to 0$ and $\nc\to 2/3$, hinting that a filled flat band has significant logarithmic corrections, see below.


\subsubsection{$n_c = \frac{2}{3}$}

The case of a filled flat band is special and needs to be treated separately. The linear term diverges meaning that a qualitatively different solution holds as $T\to 0$. Using the fact that $\mu_0(T\to 0) = 0^+$ and that the divergence from above implies that $e^{\mu_0/T}$ is large, we expand Eq.~\eqref{eq:filling_mu} in powers of $e^{-\mu_0/T}$ and keep the leading-order terms to get
\begin{align} \label{eq:mu_filled_fb}
  \mu_0/t = \frac{2\pi}{\sqrt{3}}e^{-\mu_0/T} .
\end{align}
The last equation is solved by the ansatz  $\mu_0(T) = \alpha T\ln(T_0/T)$ in the $T\to 0$ limit with $\alpha = 1$ and $T_0/t \approx 3.126$. We observe that this expression is numerically accurate only for temperatures $T/t < 10^{-4}$, significantly lower than the numerically accessible Kondo temperatures.

\subsubsection{$\frac{2}{3} < n_c < 2$}

For fillings corresponding to the dispersive part of the DOS, the low-temperature result for the chemical potential is simply
\begin{align}\label{eq:mu0_dispersive}
 \mu_0 = \frac{1}{2\tilde \rho} \left ( n_c - \frac{2}{3} \right ),
\end{align}
we note that there is no subleading term in the Sommerfeld expansion within the approximation \eqref{eq:dos_approx} of a featureless DOS. Obviously, this approximation does not cover the behavior at the Dirac and van Hove points.


\subsection{$c$ and $f$ chemical potentials at $T=0$ at weak coupling}

The hybridization between the $c-$ and $f-$ fermions renormalizes the density of states. Making use of Luttinger's theorem that states that the volume of the Fermi surface increases so as to include the $f-$electrons as well, we can get an exact expression for the interacting DOS,
\begin{align}\label{eq:interacting_DOS}
    \rho(\epsilon) = \left ( 1 + \frac{r^2}{(\epsilon + \mu + \lambda')^2} \right )\rho_0\left ( \epsilon + \mu - \frac{r^2}{\epsilon + \mu + \lambda'}\right )
\end{align}
where $\rho_0$ is the non-interacting DOS with bandwidth $D$.

To obtain the analytic expressions for the mean-field parameter $r$ and therefore the coherence temperature, it is necessary to understand how the chemical potentials $\mu$ and $\lambda'$ behave in the weak-coupling limit. This part follows closely the derivation in Ref.~\onlinecite{burdin01} and is included for completeness. Adding the two filling-related mean-field equations at $T=0$ yields
\begin{align}\label{eq:luttingers_thm}
 \frac{n_c}{2} + \frac{1}{2} = \int_{\epsilon_L}^0 d\epsilon\ \rho(\epsilon),
\end{align}
with $\rho(\epsilon)$ the interacting DOS given in Eq.~\eqref{eq:interacting_DOS} and $\epsilon_L$ is the lowest cutoff given by, $\epsilon_L =  -\mu - \frac{\lambda'}{2} - \sqrt{\left( \frac{\lambda'}{2}\right)^2 + r^2 }$. Eq.~\eqref{eq:luttingers_thm} is a manifestation of Luttinger's theorem which states in words that exactly $n_c$ $c-$ and exactly one $f-$ electron (per spin flavor) participate in the Fermi sea. Eq.~\eqref{eq:luttingers_thm} can be rewritten only in terms of the non-interacting DOS, so that it provides an alternative interpretation: the hybridization between the $c-$ and $f-$ electrons shift the chemical potential exactly by $\Delta \mu = \mu_0(n_c + 1) - \mu = -r^2/(\mu + \lambda')$.

\begin{figure}[tb]
 \centering
  \includegraphics[width=\columnwidth]{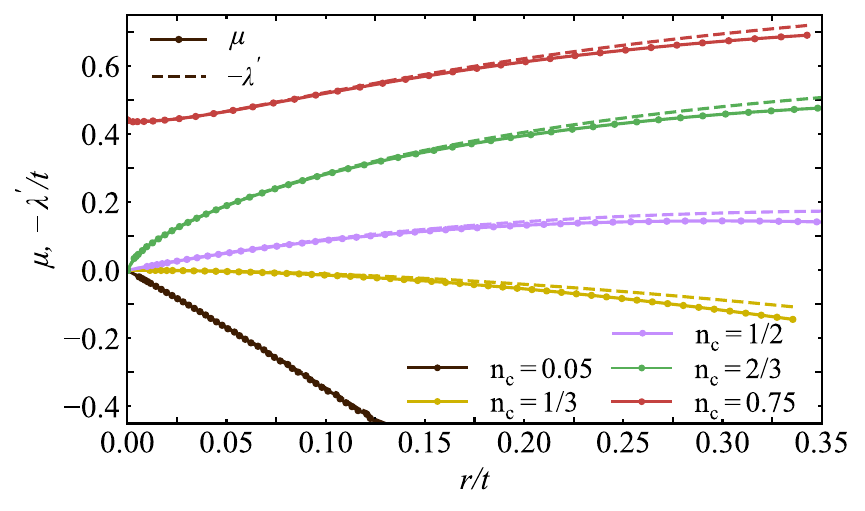}
 \caption{
 $r$ dependence of the chemical potentials $\mu$ and $-\lambda'$ as a function, which verifies their leading order behavior with respect to each other and $r$, for a series of fillings $\nc$.
}
 \label{fig:mu_lamda_chem_pots}
\end{figure}

At weak couplings it is the case that $\mu \approx \mu_0(n_c)$, so the chemical potential is unaffected, to at least first order in $r$. Given also the fact that $\Delta \mu \sim \mathcal{O}(D)$, we can readily relate the two chemical potentials via,
\begin{align}\label{eq:chem_pots_relation}
 \lambda' = -\mu + \mathcal{O}\left ( \frac{r^2}{D} \right )
\end{align}
As this derivation does not rely on the DOS being featureless, Eq.~\eqref{eq:chem_pots_relation} is expected to hold for flat-band fillings as well. Fig~\ref{fig:mu_lamda_chem_pots} verifies the relation obtained between the two chemical potentials.


\begin{figure}[tb!]
\includegraphics[width=\columnwidth]{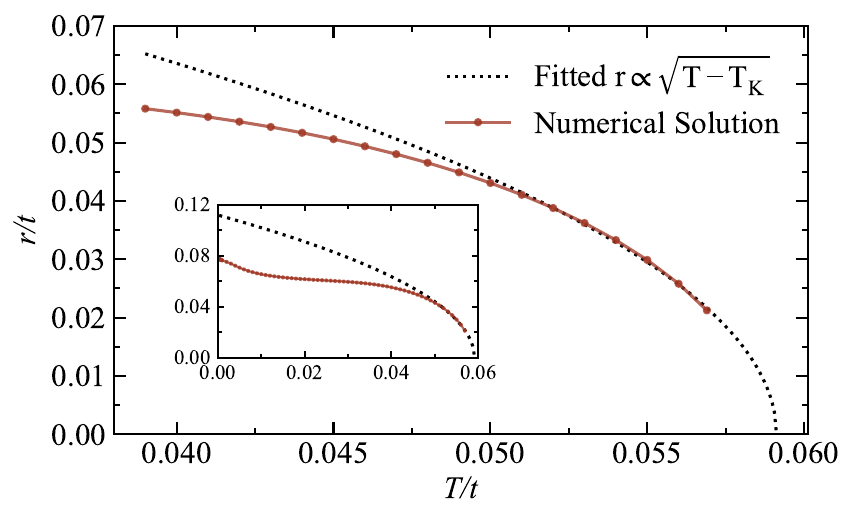}
\caption{
$T_K$ is extracted by fitting the numerical data for $r(T)$ to $r = A \sqrt{T - \TK}$ near $\TK$. Here we show this fit for $\nc = 0.03$ and $\JK = 0.97t$; the inset shows $r(T)$ down to $T=0$.
}
\label{fig:r_scaling}
\end{figure}
\begin{figure*}
  \includegraphics[width=0.95\textwidth]{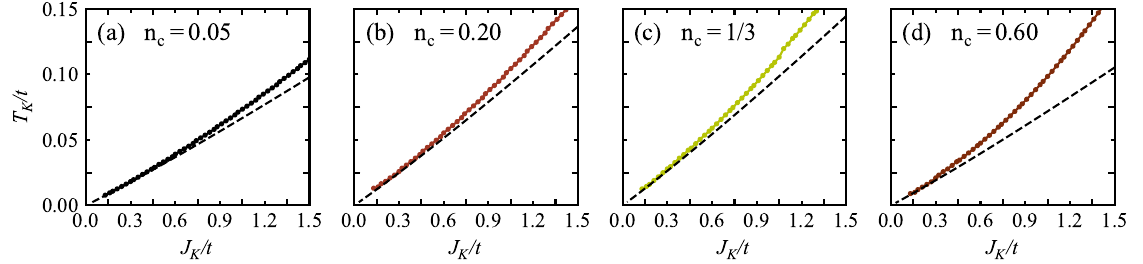}
  \caption{Numerical data (solid line) for the Kondo temperature $\TK$ with fittings of the analytical result (dashed line) of Eq.~\eqref{eq:TK_partially_filled_full} for fillings $0 < n_c < 2/3$.
  }
  \label{fig:kondo_fits_n_in_FB}
\end{figure*}
\begin{figure*}[tb]
  \includegraphics[width=0.95\textwidth]{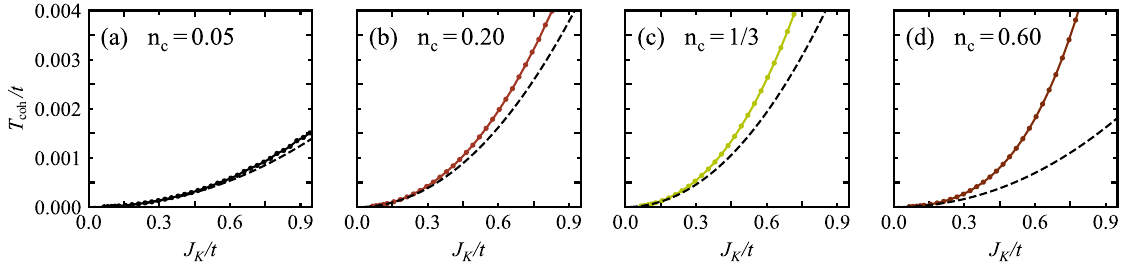}
  \caption{
  Numerical data (solid line) for the coherence temperature $\Tcoh$ with fittings of the analytical result (dashed line) of Eq.~\eqref{eq:r_vs_Jk_analytic} for fillings $0 < \nc < 2/3$.
  }
  \label{fig:coh_fits_n_in_FB}
\end{figure*}

\section{Kondo temperature}\label{apx:TKondo_solutions}

As explained above, the solution for $\TK$ follows from the single-impurity mean-field equations which lead to Eq.~\eqref{eq:kondo_final_MF}. Under the approximation taken for the DOS, the flat-band contribution is isolated from the dispersive bands contribution. The equation can then be manipulated to give,
\begin{align}\label{eq:kondo_JK_vs_TK}
    \frac{2}{J_{\rm K}} \approx \frac{1}{3}\frac{\tanh\left (\frac{\muk}{2\TK}\right )}{\muk} + \tilde \rho\ln\left( \frac{\muk(\tilde D-\muk)}{\TK^2}\right ) +   \kappa\ \tilde \rho
\end{align}
where $\kappa = 0.25127$ is a constant contribution to the integral of Eq.~\eqref{eq:kondo_final_MF} and $\muk \equiv \mu_0(\TK)$ is the chemical potential at the Kondo temperature. The last two terms are the asymptotic limit of the integral when $\TK \ll D$, so the equation is strictly valid only in the weak-coupling limit.

As explained in the main text, to numerically extract the Kondo temperature we use the fact that the order-parameter critical exponent of mean-field is $\beta = 1/2$, i.e., we expect $r \sim \sqrt{T - T_K}$. We hence fit $r(T)$ to such a dependence to obtain $\TK$, with an example displayed in Fig.~\ref{fig:r_scaling}.

\subsection{$0 < n_c < \frac{2}{3}$}
In the case of a partially fillled flat band, the chemical potential from Eq.~\eqref{eq:mu0_partially_filled} simplifies the equation but it must still be inverted for $\TK$. The latter is achieved by making the ansatz $\TK = f_0 \JK + f_n\JK^2 \ln(\JK)$, and expanding perturbatively to obtain,
\begin{align} \label{eq:TK_partially_filled_full}
    \TK = \frac{1}{2}m_{n_c} \JK - \frac{2}{3} m_{n_c} \frac{\JK^2}{4\tilde D}\ln(\JK/2\tilde D)
\end{align}
with $m_{n_c} = \frac{1}{3} \frac{\tanh(\alpha(n_c)/2)}{\alpha(n_c)}$. The comparison to numerical data is in Fig.~\ref{fig:kondo_fits_n_in_FB}, with excellent agreement. 

\subsection{$n_c = \frac{2}{3}$}
For a fully filled flat band, the form of the chemical potential is different and should thus be treated by its own. Both terms in Eq.~\eqref{eq:kondo_JK_vs_TK} may be expanded perturbatively for small $\TK$, so that the equation to be solved is,
\begin{align}
 \frac{2}{\JK} \approx \frac{1}{3}\frac{1}{\TK \ln(T_0/\TK)} \left (1 - 2\frac{\TK}{T_0} \right ) + \tilde \rho\ln\left( \frac{\ln(\TK)}{\TK}\right ) .
\end{align}
A truncation to leading order in $\TK$ allows us to solve for $\TK$ explicitly,
\begin{align} \label{eq:TK_filled_full}
    \TK = -\frac{\JK}{6} \frac{1}{W_{-1}(-\JK/6T_0)}
\end{align}
where $W_{-1}(x)$ is the lower real branch of the Lambert $W$ function. The asymptotic form of $W_{-1}(x) \approx \ln(-x) - \ln(-\ln(x))$ allows us to approximate $\TK$ as
\begin{align} \label{eq:TK_filled_full}
    \TK = \frac{\JK}{6} \frac{1}{\ln(6T_0/\JK)} \left(1-\frac{\ln\ln(6T_0/\JK)}{\ln(6T_0/\JK)}\right)\,.
\end{align}

\subsection{$0 < n_c < \frac{2}{3}$}
For fillings in the dispersive band, we should expect conventional Kondo effect behavior. This is indeed the case, since from Eq.~\eqref{eq:mu0_dispersive}, it follows that the first term in Eq.~\eqref{eq:kondo_JK_vs_TK} is a constant, so that
\begin{align} \label{eq:TK_dispersive_band}
    \TK = F_K(n_c) \exp \left (-\frac{1}{\tilde \rho \JK} \right )
\end{align}
up to a filling-dependent prefactor. Note that $\tilde \rho$, a constant, replaces the conventional $\rho_0(n_c)$ as a result of the approximation to the DOS.


\section{Coherence temperature solution}\label{apx:Tcoh_solutions}

We present in this sections some remarks on how Eq.~\eqref{eq:MF_1}-\eqref{eq:MF_3} is solved, in the weak-coupling limit and as $T\to 0$.
The approximate form of the DOS, Eq.~\eqref{eq:dos_approx}, and the relation between $\mu$ and $\lambda'$, allow us to write the mean-field equation for $\JK$ in the form,
\begin{align}\label{eq:coherence_MF_final}
\frac{1}{\JK} = \frac{1}{3} \frac{1}{\sqrt{\lambda'^2/4 + r^2}} - 2\tilde \rho \ln \Big [ \frac{r^2  \left( \lambda'/2 + \sqrt{\lambda'^2/4 + r^2} \right ) }{\Delta \mu}\Big ]
\end{align}
where $\Delta \mu = \mu_0(n_c + 1) - \mu = \mathcal{O}(D)$. The chemical potentials can be determined for each $n_c$ by the second mean-field equation,
\begin{align} \label{eq:rw_filling_eq}
n_c = \frac{1}{3} \left [ 1 - \frac{\lambda'/2}{\sqrt{\frac{\lambda'^2}{4} + r^2}}\right ] - 2\tilde \rho \left [ \frac{\lambda'}{2} - \sqrt{\frac{\lambda'^2}{4} + r^2}\ \right ]
\end{align}

Eq.~\eqref{eq:rw_filling_eq} allows us to solve for $\lambda' = f(r)$ for each filling, to leading order in each case,
\begin{align}
\begin{aligned}[c]
0<n_c&<1/3:\\
n_c &= 1/3:\\
1/3<n_c&<2/3:\\
n_c &= 2/3:\\
2/3<n_c&<2:
\end{aligned}
\hspace{1cm}
\begin{aligned}[c]
\lambda' &\approx g_n r\\
\lambda' &\approx 12\tilde \rho r^2\\
\lambda' &\approx g_n r\\
\lambda' &\approx g_{2/3} (r^{2}t)^{1/3}\\
\lambda' &\approx \text{const.}
\end{aligned}
\end{align}
where $g_n = \frac{2(\frac{1}{3} - n)}{\sqrt{n(2/3 - n)}}$ for $n \neq \frac{1}{3}, \frac{2}{3}$ and $g_{2/3} = -(2\pi/\sqrt{3})^{1/3}$.
For each filling, we can substitute $\lambda' = f(r)$ in Eq.~\eqref{eq:coherence_MF_final} and invert the equation perturbatively in $\JK$. We quote the results here,
\begin{align} \label{eq:r_vs_Jk_analytic}
\begin{aligned}[c]
0<n_c <2/3&:\\
n_c = 2/3&:\\
2/3<n_c<2&:
\end{aligned}
\hspace{1cm}
\begin{aligned}[c]
r &= h_n \JK - 3\tilde \rho h_n \JK^2 \ln(\JK)\\
r &= h_{2/3} \JK^{3/2}/t^{1/2}\\
r &= h_n \exp\left (-\frac{1}{2\tilde \rho \JK} \right )
\end{aligned}
\end{align}
where $h_{n} = \frac{1}{2}\sqrt{n(\frac{2}{3} - n)}$ for $n\neq \frac{1}{3},\frac{2}{3}$, $h_{1/3} = 1/6$, $h_{2/3} = 1/(3^{5/4}\sqrt{2\pi})$ and $h_{n>2/3}=f(n_c)$.

The fittings of this results to the numerical data are depicted in Fig.~\ref{fig:coh_fits_n_in_FB} for $0<n_c<\frac{2}{3}$; the data for $n_c = \frac{2}{3}$ are included in Fig.~\ref{fig:logTcoh_logJk}. For completeness, we note that the deviations from the asymptotic limit are in agreement with the expectation that Kondo physics is largely governed by crossover effects away from weak coupling.

We note that the agreement between the numerical data at the special fillings  $n_c = 1/6, 1/3, 1/2$, for which the chemical potential in the hybridized system lies on a vHs or a Dirac point, and the analytical solution using a flat DOS indicate that the mean-field parameters are essentially insensitive to the features of the hybridized band structure (as opposed to that of the bare $c$ electrons)

\begin{figure}[!bt]
\centering
\includegraphics[width=\columnwidth]{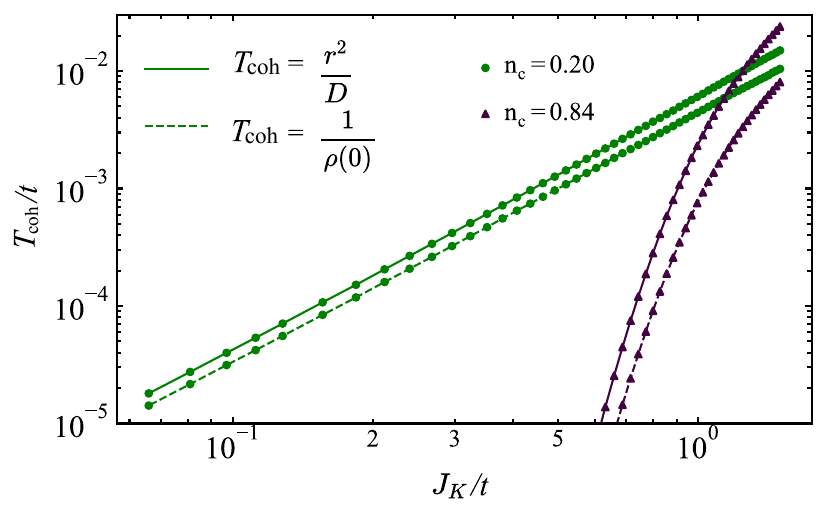}
\caption{
Comparison of the mean-field $\Tcoh$, calculated either as $r^2(T=0)/D$ (solid line) or as $1/\rho(0)$ (dashed line), for $\nc = 0.2$ ($\nc=0.84$) in the flat (dispersive) band, respectively.
}
\label{fig:comparison_tcoh}
\end{figure}

We finally provide a comparison between our simple definition of the coherence temperature, $\Tcoh=r^2/D$, to that of $\Tcoh = \frac{1}{\rho(\w\!=\!0)}$ which relates $\Tcoh$ to the bandwidth of the emergent heavy fermions. Both are strictly equivalent only for a featureless conduction-electron DOS. A direct comparison of the two expressions using numerical data is in Fig.~\ref{fig:comparison_tcoh}, which shows that same qualitative behavior of both, but a difference in numerical prefactors. Obviously, this agreement is not present at $n_c = 1/6, 1/3, 1/2$ where $\rho(\w\!=\!0)$ is either zero or infinite.


\section{Kondo temperature at large $\JK$}

As an isolated flat band has vanishing kinetic energy, it is instructive to consider a Kondo lattice in the (somewhat artificial) limit of $\JK \gg t$. Here we expect that also $T_K \gg t$. We can think of a system of dimers, each formed by one $c$ and one $f$ site, which are weakly coupled by $t$. At half-filling, i.e. $\nc=1$, each dimer hosts two electrons forming a singlet, and we expect $\TK \propto \JK$.

Indeed, we can solve Eq.~\eqref{eq:filling_kondo_mu} for the case of a single narrow band in the limit $T\gg t$, with the result $\mu(T) = \beta(n_c) T$ with, $\beta(n_c) = \ln \left ( \frac{n_c}{2-n_c}\right )$. Then, the large-$\JK$ solution of the Kondo equation is
\begin{align}
    \TK = \frac{1}{2\beta(n_c)} \tanh \left ( \frac{\beta(n_c)}{2}\right )\JK .
\end{align}
being symmetric about $n_c = 1$, as expected from a picture of isolated dimers.

We note that Fermi-liquid coherence in a Kondo lattice with $\JK \gg t$ is limited by $t$ (not by $\JK$), as the effective carriers display a hopping rate $t$. As a result, we will have $\Tcoh\propto t$. Similarly, any effects of inter-moment interaction are limited by $t$, such that Kondo screening is realized in the limit $\JK \gg t$ also beyond mean-field theory. This does not exclude that the (heavy-fermion) metal emerging for $\nc\neq 1$ is unstable against symmetry-breaking order.



\begin{thebibliography}{99}
\bibitem{tsignnote}
In the Hamiltonian \eqref{eq:KLM_hamiltonian} we have chosen a positive sign for the hopping such that the flat band appears at the bottom of the single-particle spectrum in order to simplify notations. The physics for the opposite-sign case can be simply obtained by a particle--hole transformation.
\bibitem{tknote}
Note that the mean-field theory for a model of SU(2) electrons coupled to an impurity with a degeneracy of $N=2j+1$ yields a Kondo temperature $\TK \propto \exp(\frac{2}{(N-1)\rho\JK})$ \cite{hewson}.
\bibitem{prefactor_comment}
Note that $m_{n_c}$ diverges at $\nc =1/6$ but its limit as $\nc\to 1/6$ is finite.
\end{thebibliography}
\end{document}